\documentclass[aps,prl,twocolumn,showpacs]{revtex4}

\usepackage{graphicx}
\usepackage{amssymb}
\usepackage{hyperref}

\begin{document} 

\title{Scattering-free plasmonic optics with anisotropic metamaterials}

\author{Justin Elser}
\author{Viktor A. Podolskiy}
\affiliation{Physics Department, Oregon State University, 301 Weniger Hall, Corvallis, OR 97331, USA}

\pacs{71.36.+c, 42.25.Ja, 42.25.Lc}

\date{\today}

\begin{abstract}
We develop an approach to utilize anisotropic metamaterials to solve one of the fundamental problems of modern plasmonics -- parasitic scattering of surface waves into free-space modes, opening the road to truly two-dimensional plasmonic optics. We illustrate the developed formalism on examples of plasmonic refractor and plasmonic crystal, and discuss limitations of the developed technique and its possible applications for sensing and imaging structures, high-performance mode couplers, optical cloaking structures, and dynamically reconfigurable electro-plasmonic circuits. 
\end{abstract}

\maketitle

An interface between two materials with opposite signs of dielectric permittivity, such as that between a metal and dielectric, can support a highly-confined surface electromagnetic wave, known as a surface plasmon polariton (SPP)\cite{raetner}. SPPs are the enabling mechanism for sub-diffraction sensing, imaging, and subwavelength light guiding\cite{plasmonFormer}. These applications are ultimately unified in the paradigm of surface optics -- where surface waves -- rather than plane waves -- are used for on-chip optical communications between nm-sized ports. While a number of surface optical elements, ranging from waveguides, to lenses, to reflectors\cite{plasmonFormer} have been already designed, the performance of the majority of these devices is fundamentally flawed by the parasitic scattering of SPPs from the boundaries between optical elements with different refractive indices. Typically, 10\ldots30\% of SPP energy scatters into free-space modes at a single boundary\cite{zhangPRB}, severely hindering the performance of surface optical elements and essentially making it impossible to realize the 2D optics paradigm with existing isotropic materials. Here we demonstrate that properly designed anisotropic metamaterials can be utilized to completely eliminate this parasitic scattering by decoupling the response of plasmonic circuits to different polarizations of electromagnetic radiation, and thus opening the roadway to {\it truly plasmonic optics}. We further demonstrate that the optical properties of anisotropic plasmonic circuits can be dynamically modulated with external electric fields. Finally, we discuss the implications of polarization decoupling to other applications of anisotropic metamaterials, including negative refraction, sub-diffraction imaging, and cloaking\cite{NIM,cloak,hyperlens}. 

\begin{figure}[t]
\centerline{\includegraphics[width=7.5cm]{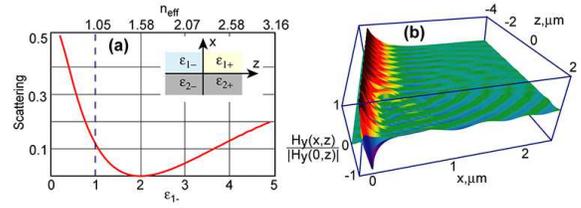}}
\caption{\label{figIso} 
(color online) (a) Fraction of energy of incident SPP scattered into plane waves from the interface between two SPP-supporting media with different SPP refractive indices; $\epsilon_{1+}=2; \epsilon_{2+}=\epsilon_{2-}=-10$; (b) normalized field structure corresponding to $\epsilon_{1-}=1$ in (a)
} 
\end{figure}

The SPP is a solution of Maxwell equations that represents a transverse-magetic (TM) wave propagating at the interface between two materials. The field of an SPP has harmonic in-plane structure [$\propto \exp(-i\omega t+ i k_y y+i k_z z)$] and exhibits exponential decay [$\propto \exp(-\kappa_j |x|)$] away from the interface. Its spatial behavior can be related to permittivities of materials $\epsilon_1$ and $\epsilon_2$, angular frequency $\omega$, and speed of light in the vacuum $c$, via: \cite{raetner} 
\begin{equation}
\label{eqSPPiso}
K^2=k_y^2+k_z^2=\frac{\omega^2}{c^2}
\frac{\epsilon_1 \epsilon_2}{\epsilon_1+\epsilon_2}, \;
\kappa_j^2=K^2-\epsilon_j\frac{\omega^2}{c^2}, 
\end{equation}
with $j=1,2$ corresponding to top and bottom materials (see Fig.\ref{figIso}). The ratio $n_{\rm eff}=K c/\omega$ is known as the modal refractive index of the SPP.

As seen from Eqs.(\ref{eqSPPiso}), propagation of SPPs can be controlled by varying the optical properties of either of two materials. In practice, such a variation is achieved by deposition of high-index dielectric on top of metal (changing $\epsilon_1$), changing metallic substrate (changing $\epsilon_2$), or corrugation of the interface \cite{plasmonFormer,noeckelScat}. Unfortunately, a variation of $n_{\rm eff}$ necessarily leads to change of the spatial profile of this mode (given by parameters $\kappa_1$ and $\kappa_2$). As a result, reflection and refraction of surface waves is fundamentally different from those of plane waves. 

Plane waves in a sense form a closed space -- a single {\it plane wave} incident on the interface between two isotropic media with different refractive indices {\it excites a set of two plane waves}: one reflected wave, and one transmitted wave. In contrast to this behavior, refraction of SPPs is accompanied by the parasitic out-of-plane scattering; a single SPP incident at the boundary between two surface elements with different $n_{\rm eff}$ excites a set of scattered plane waves in addition to reflected and transmitted SPPs (Fig.\ref{figIso})\cite{zhangPRB}. In typical plasmonic optics, $\sim 20\%$ of energy is scattered into plane waves in individual reflection from the boundary between two elements. The implications of this out-of-plane scattering are not limited to dramatic reduction of SPP intensity. In fact, the scattering provides a mechanism for {\it coupling} between surface and plane waves, and thus it creates a possibility for plane waves to couple back into surface modes and {\it temper the integrity of surface-mode signals}. 

The parasitic scattering is further increased in oblique refraction when TM-polarized SPPs couple to both TM and TE plane wave spectrum.
 
The out-of-plane scattering of SPPs into free-space modes can be eliminated by meeting two conditions. First, the spatial profile of the SPP mode should be independent of its refractive index. And second, the boundary between the optical elements should not support inter-polarization (TE $\leftrightarrow$ TM) coupling.

The purpose of this work is to show that both these conditions can be satisfied in uniaxial anisotropic media, to derive the description of optical properties of SPP in anisotropic structures, and to illustrate the developed formalism on the examples of realistic metamaterials. 

We start by deriving dispersion equations for the SPP propagating at the interface between two anisotropic structures with optical axes perpendicular to the interface via standard wave-matching approach: 
\begin{equation}
\label{eqSPPaniz}
K^2=\frac{\omega^2}{c^2}
\frac{\epsilon_1^x \epsilon_2^x(\epsilon_1^{yz}-\epsilon_2^{yz})}
{\epsilon_1^x\epsilon_1^{yz}- \epsilon_2^x \epsilon_2^{yz}}, \; 
\kappa^2_j=\epsilon_j^{yz}
\left(
\frac{K^2}{\epsilon_j^x}-\frac{\omega^2}{c^2}
\right)
\end{equation}
(these equations replace Eqs.(\ref{eqSPPiso}) in anisotropic media). Note that the SPPs propagate perpendicular to optical axes of the media, and therefore their propagation is completely isotropic despite material anisotropy. 

A fundamental advantage of anisotropic media over their isotropic counterparts lies in the ability to {\it independently control} propagation parameter of the SPP $K$ and its structural profile by changing individual components of the permittivity tensor. In particular, it becomes possible to select the permittivities of anisotropic media so that the SPP profile becomes independent of the modal index, . Then mismatch of SPP modes in different optical elements is eliminated and parasitic scattering of SPPs is vanished leading to {\it purely 2D optics}. 

\begin{figure}[t]
\centerline{\includegraphics[width=7.5cm]{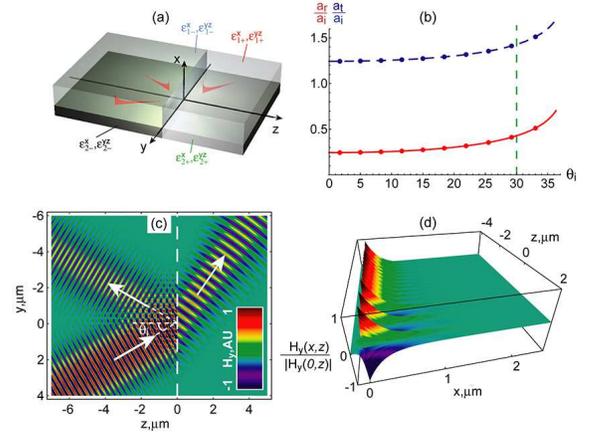}}
\caption{\label{figRef} (color online) (a) Truly plasmonic optics: no free-space modes are excited in SPP refraction; (b) Fresnel relations for transmitted (blue, dashed) and reflected (red, solid) SPPs. Dots and lines correspond to numerical solution of Maxwell equations and analytical Eqs.(\ref{eqFresnelR}) respectively; $\epsilon^{x}_{1-}=2.7; \epsilon^{yz}_{1-}=\epsilon^{yz}_{1+}=\epsilon^x_{1+}=1; \epsilon^x_{2-}=\epsilon^{yz}_{2-}=\epsilon^{yz}_{2+}=-10; \epsilon^x_{2+}=-3.7;
$ (c) refraction of SPPs; arrows correspond to Eq.(\ref{eqSnell}); $H_y$ component of the field across $x=50 nm$ plane is shown; (d) normalized field distribution along $y=0$ axis. Note that SPPs have constant $x$-profile despite change in refractive index (wavelength) at $z=0$ (Compare to Fig.\ref{figIso}b).  
} 
\end{figure}

To derive the relationship between components of permittivity tensor to realize surface optics, it suffices to consider a refraction of an SPP through the boundary between two SPP-supporting structures. Denoting the properties of materials at the left side of the boundary with ``$-$'' sign, those at the right side of the boundary with ``$+$'' sign (see Fig.\ref{figRef}), and requiring that the profile of SPP mode is unchanged across the interface:
\begin{eqnarray}
\label{eqCont1}
\kappa^2_{1-}=\kappa^2_{1+}, \;
\kappa^2_{2-}=\kappa^2_{2+},
\end{eqnarray}
we obtain
\begin{eqnarray}
\label{eqCont2}
\frac{\epsilon^{x}_{1+}}{\epsilon^x_{1-}}=
\frac{\epsilon^{x}_{2+}}{\epsilon^x_{2-}}, \;
\epsilon^{yz}_{1-}= \epsilon^{yz}_{1+}, \;
\epsilon^{yz}_{2-}= \epsilon^{yz}_{2+}, 
\end{eqnarray}
where the ratio of $x$-components of permittivity tensors controls the change of modal index ($n_{\rm eff}=K c/\omega$): 
\begin{equation}
\label{eqContN}
\frac{n^2_{+}}{ n^2_{-}}=
\frac{K^2_{+}}{ K^2_{-}}=
\frac{\epsilon^{x}_{1+}}{\epsilon^x_{1-}}.
\end{equation}

An ideal surface optical system therefore has constant in-plane ($\epsilon^{yz}$) components of the permittivity tensors and only modulates the permittivities along the optical axis ($\epsilon^x$). The structure becomes completely transparent to TE-polarized radiation. Hence, the TE waves do not scatter from interfaces and do not couple to any TM waves. The {\it independent of refractive index} $x$-profile of SPPs further prevents coupling between surface modes and TM-polarized volume waves.

When Eqs.(\ref{eqCont2}) are satisfied, the behavior of surface waves in surface optical circuits can be mapped to the familiar laws of 3D optics. This way, when the SPP undergoes the refraction through the boundary between two surface optical elements, the directions of reflected and refracted beams are related to the direction of incident beam through Snell's law, and the amplitudes of the $E^x_1$ components of refracted and reflected beams ($a_t, a_r$) are related to the amplitude of $E^x_1$ component of incident SPP ($a_i$) through Fresnel equations identical to those in 3D optics:
\begin{eqnarray}
\label{eqSnell}
\frac{\sin(\theta_i)}{n_-}=\frac{\sin(\theta_r)}{n_{-}}
=\frac{\sin(\theta_t)}{n_+}
\\
\label{eqFresnelR}
\frac{a_r}{a_i}=\frac{k_{z-}-k_{z+}}{k_{z-}+k_{z+}}, \;
\label{eqFresnelT}
\frac{a_t}{a_i}=\frac{2k_{z-}}{k_{z-}+k_{z+}}
\end{eqnarray}
The perfect agreement between analytical Eqs.(\ref{eqSnell},\ref{eqFresnelR}) and numerical solutions of Maxwell equations in realistic non-scattering plasmonic structures is shown in Fig.\ref{figRef}.

\begin{figure}[t]
\centerline{\includegraphics[width=7.5cm]{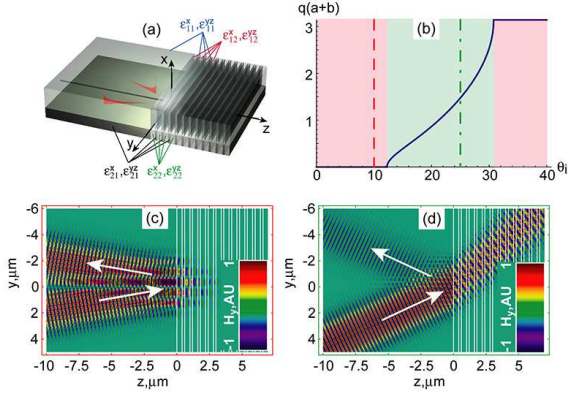}}
\caption{\label{figPC} (color online) (a) Geometry of the plasmonic Bragg reflector (b) the dependence of Bragg vector on the SPP incidence angle at $\lambda=500 nm$; dashed and dash-dotted lines correspond to panels (c) and (d); (c) the regime of ``surface mirror''; $H_y$ component of the field at $x=50nm$. (d) transparency band of the same plasmonic crystal; 
$\epsilon^x_{11}=2.7; \epsilon^{yz}_{11}=\epsilon^{yz}_{12}=\epsilon^x_{12}=1; \epsilon^x_{21}=\epsilon^{yz}_{21}=\epsilon^{yz}_{22}=-10; 
\epsilon^{x}_{22}=-3.7$   
} 
\end{figure}

The analogy between 2D and 3D optics naturally extends to SPP propagation in structures with periodically modulated refractive index. When Eqs.(\ref{eqCont2}) are satisfied so that the SPPs cannot be scattered into volume modes, periodic modulation of $\epsilon^x$ will create surface analog of photonic crystals\cite{jonnapoulos} -- non-scattering plasmonic crystals. For the simplest case of two-component layered plasmonic crystal illustrated in Fig.\ref{figPC}, the dispersion relation is identical to that of a Bragg mirror\cite{yariv}:
\begin{equation}
\cos[q (a+b)]=\cos[k_{z1} a] \cos[k_{z2} b]-
\gamma 
\sin[k_{z1} a] \sin[k_{z2} b],
\end{equation}
where $a$ and $b$ are thicknesses of SPP-supporting surface elements formed by materials with dielectric permittivities $[\{\epsilon^x_{11}, \epsilon^{yz}_{11}\}, \{\epsilon^x_{21}, \epsilon^{yz}_{21}\}]$ and $[\{\epsilon^x_{12}, \epsilon^{yz}_{12}\}, \{\epsilon^x_{22}, \epsilon^{yz}_{22}\}]$ respectively, $\gamma=\frac{1}{2}\left(\frac{k_{z1}}{k_{z2}}+\frac{k_{z2}}{k_{z1}}\right)
$, and $k_{zi}$ are $z$-components of SPP wave-vectors in these materials.

We now discuss the perspectives of experimental realization of non-scattering surface optical elements. While few natural materials exhibit required anisotropic response, efficient control over components of permittivity tensor can be achieved in metamaterials -- nanostructured composites with tailored optical properties. Thus, multilayer- or nanowire composites can be readily utilized to fabricate the anisotropic structure with arbitrary dielectric permittivities (see Refs.[\onlinecite{NIM,miltonBook}] and references therein). Here we utilize the effective-medium theory\cite{miltonBook} to assess the perspectives of employing the nanolayered composites for 2D optics.

In particular, we use a combination of two SPP-supporting structures. The first structure guides surface modes at the interface between a metamaterial with $\epsilon^x=2.7; \epsilon^{yz}=1$ and an isotropic medium with $\epsilon=-10$ resembling Ag-silica composite and Ag respectively at vacuum wavelength $\lambda_0=500 nm$. The second structure supports an SPP at the interface between vacuum and a metamaterial with $\epsilon^x=-3.71; \epsilon^{yz}=-10$ corresponding to Al-Au multilayer \cite{palik}. 

The absence of SPP scattering into propagating modes regardless of incident angle is evident from ideal field matching across the systems and from the position-independent mode profile (see Figs.\ref{figRef},\ref{figPC}). 

\begin{figure}[t]
\centerline{\includegraphics[width=7.5cm]{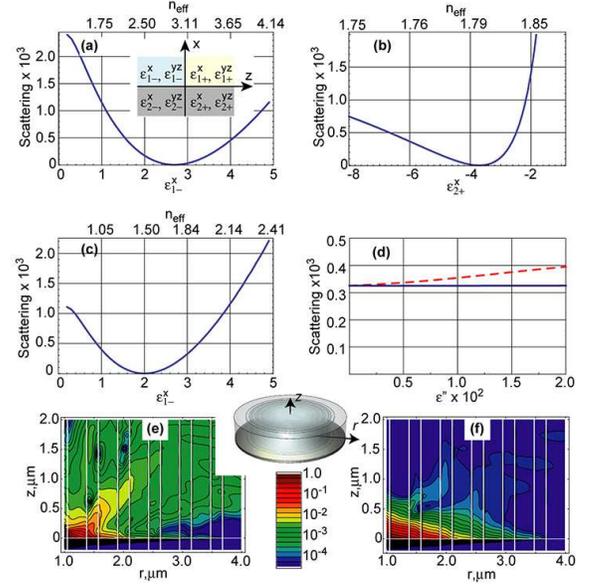}}
\caption{\label{figTol} (color online) Scattering at the interface between two SPP structures (a) $\epsilon^x_{2+}=-3.71$, $\epsilon^x_{1-}$ is varied; (b) $\epsilon^x_{1-}=2.7$, $\epsilon^x_{2+}$ is varied; remaining parameters in (a,b):  $\epsilon^{yz}_{1-}=\epsilon^x_{1+}=\epsilon^{yz}_{1+}=1; \epsilon_2=\epsilon^x_{2-}=\epsilon^{yz}_{2-}=\epsilon^{yz}_{2+}=-10$; 
(c) scattering in the system with common metal substrate $\epsilon^{yz}_{1-}=\epsilon^x_{1+}=\epsilon^{yz}_{1+}=2; \epsilon^x_{2-}=\epsilon^{yz}_{2-}=\epsilon^x_{2+}=\epsilon^{yz}_{2+}=-10$ as a function of $\epsilon^x_{1-}$; (d) effect of losses in the structure in (c) on scattering; Red dashed line: $\epsilon^x_{1-}=3(1+i\epsilon '')$; $\epsilon^{yz}_{1-}=2(1+i\epsilon '')$; Blue  solid line: $\epsilon_2=-10(1-i\epsilon'')$. (e) scattering of $m=0$ TM mode in layered structure proposed in Ref.[\onlinecite{smolyaninovHyper}] (inset); normalized energy flux is shown; $\epsilon_2=-3.5+2.8i; \epsilon_{\rm PMMA}=2.25; \lambda_0=0.51\mu m$; (f) same as (e), but the polymer is anisotropic with $\epsilon^z=2.62,\epsilon^{r\phi}=1$. 
} 
\end{figure}

To analyze the limitations of non-scattering surface optics and its tolerance to experimental imperfections, we study the parasitic scattering resulting mismatch of dielectric permittivity. The analysis is performed via numerical solutions of Maxwell equations with the commercial finite-element PDE solver, COMSOL multiphysics 3.3a. Results of these simulations are summarized in Fig.\ref{figTol}. It can be clearly seen that non-scattering plasmonics is highly tolerant to variation in $\epsilon_x$. Furthermore, when $|\epsilon_2|\gg\epsilon_1$, the sensitivity of scattering to variations of ``metallic'' ($x<0$) component of the structures is almost undetectable since the field is primarily concentrated in the dielectric. 

Such a high tolerance of the surface optics formalism with respect to variations of material permittivity allows one to realize high performance optical circuits on common metallic substrates. An example of such a system is shown in Fig.\ref{figTol}(c,d). The structure comprises several dielectrics with matched $\epsilon^{yz}$ deposited on Ag substrate. It can be clearly seen that the parasitic scattering in this system rarely exceeds 1\% -- orders of magnitude smaller than in a comparable isotropic plasmonic system since the SPP structure is primarily affected by the $yz$ component of dielectric permittivity when $|\epsilon_2|\gg\epsilon_1$. Hence, the modal mismatch in anisotropic composites can be substantially smaller than in their isotropic counterparts. 

The advantages offered by anisotropic media are further illustrated in Fig.\ref{figTol}(e,f), where we compare the performance of imaging structure suggested in Ref.[\onlinecite{smolyaninovHyper}] to its anisotropic analog. We assume that both plasmonic structures have identical isotropic homogeneous Au substrate and introduce anisotropy only to dielectric components of the systems. The scattering in anisotropic system is suppressed by two orders of magnitude.

A set of applications of the developed formalism lie in tunable plasmonic circuits with external electric control. These structures would be based upon a single metamaterial system comprising electro-optical component (for example, a metal substrate covered with an electro-optical polymer) and electric circuitry to control the properties of this component. The static electric field, directed along $x$ axis will modify $x$ component of electro-optical metamaterial\cite{landauECM}, providing a dynamical modulation of local refractive index. This way, the optical elements (lenses, mirrors, or band-gap structures) in these systems can be created and destroyed by changing the external electric field without any structural modifications. 

Decoupling between TE and TM waves offered by anisotropic media has its advantages far beyond 2D optics: in optical fiber communications and in waveguide design, independent manipulation of mode structure and its effective index can be used as an additional control mechanism to match modes of different guiding structures and to modulate structure-dependent losses; in anisotropy-based negative refraction systems\cite{NIM} it can be used to reduce scattering losses; in coordinate-transformation-based optical cloaking applications\cite{cloak} polarization separation can be used to construct a unique system that would completely decouple the optical pathways of TM and TE waves in the bulk of metamaterial structures: TM waves would travel around the cloaked region, while TE radiation will travel through this region. 

The developed formalism, although presented here on the example of single-interface surface structures, can be further generalized to suppress or eliminate scattering in multilayered systems\cite{avrutskyLayers} and in structures with curved interfaces\cite{hyperlens}. With proper choice of materials, the developed technique can be realized in different frequency ranges, including UV, optical, and far-IR systems.

This research was supported by ONR (grant \#N00014-07-1-0457) and NSF (grant \#ECCS-0724763)

\end{document}